\documentclass{article}
 \def\tfr#1#2{{\textstyle{#1\over #2}}}
 \def\ac{arbitrary constant}
 \def\isc{invariant surface condition}
 \def\afs{arbitrary functions}
 \def\inl{infinitesimal}
 \def\inls{infinitesimals}
 \def\wrt{with respect to}
 \def\deq{determining equation}
 \def\deqs{determining equations}
 \def\pde{partial differential equation}
 \def\pdes{partial differential equations}
 \def\ode{ordinary differential equation}
 \def\odes{ordinary differential equations}

\begin{document}
\title{Symmetries of a class of Nonlinear Third Order Partial Differential
Equations}
\author{P.A.~Clarkson, E.L.~Mansfield \&  T.J.~Priestley\\
        Institute of Mathematics and Statistics\\
        University of Kent at Canterbury\\
        Canterbury, CT2 7NF, U.K.\\[1.0ex]
        {\tt  P.A.Clarkson@ukc.ac.uk}\\
        {\tt  E.L.Mansfield@ukc.ac.uk}\\ 
        {\tt  tjp1@ukc.ac.uk} 
        }
\date{}
\maketitle
\begin{abstract}
In this paper we study symmetry reductions of a class
of nonlinear third order \pdes
$$ 
u_t -\epsilon u_{xxt} +2\kappa u_x= u u_{xxx} +\alpha u u_x +\beta u_x u_{xx}\,,
\eqno{(1)}
$$
where $\epsilon$, $\kappa$, $\alpha$ and $\beta$ are arbitrary constants. Three
special cases of  equation (1) have appeared in the literature,
up to some rescalings. In each case the equation
has admitted unusual travelling wave solutions: the
Fornberg-Whitham equation, for the parameters
$\epsilon=1$, $\alpha=-1$, $\beta=3$ and $\kappa=\tfr12$, admits a wave of 
greatest height, as a peaked limiting form of the travelling wave solution; 
the Rosenau-Hyman equation, for the parameters $\epsilon=0$, $\alpha=1$,
$\beta=3$ and $\kappa=0$, admits a ``compacton'' solitary wave solution; and
the Fuchssteiner-Fokas-Camassa-Holm equation, for the parameters $\epsilon=1$,
$\alpha=-3$ and $\beta=2$, has a ``peakon'' solitary wave solution.

A catalogue of symmetry reductions for equation (1) is obtained using the 
classical  Lie method and the nonclassical method due to Bluman and Cole.
\end{abstract}

\def\cite#1{[#1]}
\def\hide#1{}
\newcount\refno
\refno=0

 \def\jpa{J.\ Phys.\ A: Math.\ Gen.}
 \def\jmp{J.\ Math.\ Phys.}
 \def\NC{Nuovo Cim.}
 \def\mja{M.J. Ablowitz}
 \def\pac{P.A. Clarkson}
 \def\hs{H. Segur}
 \def\liz{E.L. Mansfield}

\def\refpp#1#2#3#4{\global\advance\refno by 1
\expandafter \xdef\csname #1\endcsname {\the\refno}\relax}{}{}{}
\def\refjl#1#2#3#4#5#6{\global\advance\refno by 1
\expandafter \xdef\csname #1\endcsname {\the\refno}\relax}{}{}{}{}{}{}
\def\refbk#1#2#3#4#5{\global\advance\refno by 1
\expandafter \xdef\csname #1\endcsname {\the\refno}\relax}{}{}{}{}
\def\refeb#1#2#3#4#5#6#7{\global\advance\refno by 1
\expandafter \xdef\csname #1\endcsname {\the\refno}\relax}{}{}{}{}{}{}
\def\reftoap#1#2#3#4#5{\global\advance\refno by 1
\expandafter \xdef\csname #1\endcsname {\the\refno}\relax}{}{}{}{}

\refbk{refAC}{\mja\ and \pac}{``Solitons, Nonlinear Evolution
Equations and Inverse Scattering''}{{\it LMS Lecture Notes in
Mathematics\/},  vol.\ {\bf 149}, Cambridge University Press, Cambridge}{1991}
\refjl{refAKNS}{M.J. Ablowitz, D.J. Kaup, A.C. Newell and H. Segur}{Stud. Appl.
Math.}{53}{249--315}{1974}
\refjl{refARS}{\mja, A. Ramani and \hs}{Phys. Rev.
Lett.}{23}{333--338}{1978}
\refjl{refARSi}{\mja, A. Ramani and \hs}{\jmp}{21}{715--721}{1980}
\refjl{refA}{W.F. Ames}{Appl. Num. Math.}{10}{235--259}{1992}
 \refjl{refABH}{D. Arrigo, P. Broadbridge and J.M. Hill}{\jmp}{34}{4692
--4703}{1993}
\refjl{refBBM}{T.B. Benjamin, J.L. Bona and J. Mahoney}{Phil. Trans. R.
Soc.  Lond. Ser. A}{272}{47--78}{1972}
\refjl{refBC}{G.W. Bluman and J.D. Cole}{J. Math. 
Mech.}{18}{1025--1042}{1969}
\refjl{refCH}{R. Camassa and D.D. Holm}{Phys.
Rev. Lett.}{71}{1661--1664}{1993}
\refjl{refCHH}{R. Camassa, D.D. Holm and J.M. Hyman}{Adv. Appl.
Mech.}{31}{1--33}{1994}
\refjl{refCHW}{B. Champagne, W. Hereman and P. Winternitz}{Comp.
Phys. Comm.}{66}{319--340}{1991}
\refjl{refPACiii}{P.A. Clarkson}{Europ. J. Appl. Math.}{1}{279--300}{1990}
\reftoap{refPAC}{\pac}{{\rm ``Nonclassical Symmetry Reductions of the Boussinesq
Equation'',}\ Chaos, Solitons and Fractals}{to appear}{1995}
\refjl{refCFA}{\pac, A.S. Fokas and M.J. Ablowitz}{SIAM J. Appl.
Math.}{49}{1188--1209}{1989}
\refjl{refCK}{\pac\ and M.D. Kruskal}{J. Math. Phys.}{30}{2201--2213}{1989}
\refjl{refCMi}{\pac\ and \liz}{Physica}{D70}{250--288}{1994}
\refjl{refCMii}{\pac\ and \liz}{Nonlinearity}{7}{975--1000}{1994}
\refjl{refCMiii}{\pac\ and \liz}{SIAM J. Appl. Math.}{54}{1693--1719}{1994}
\refjl{refCMiv}{\pac\ and \liz}{Acta Appl. Math.}{39}{245--276}{1995}
\refjl{refCFP}{R. Conte, A.P. Fordy and A.
Pickering}{Physica}{D69}{33--58}{1993}
\refjl{refCS}{F. Cooper and H. Shepard}{Phys.
Lett.}{A194}{246--250}{1994}
\refpp{refFoka}{A.S. Fokas}{``Moderately long water waves of
moderable amplitude are integrable'', preprint}{(1994)}
\refpp{refFokb}{A.S. Fokas}{``A new class of physically important integrable
equations'', preprint, Department of Mathematics and Computer Science, Clarkson
University, Potsdam, NY}{(1994)}
\refpp{refFokS}{A.S. Fokas and P.M. Santini}{``An inverse acoustic
problem and linearization of moderate amplitude dispersive waves'',
preprint}{(1994)}
\refjl{refFuch}{B. Fuchssteiner}{Progr. Theor. Phys.}{65}{861--876}{1981}
\refpp{refFuchb}{B. Fuchssteiner}{``Some tricks from the symmetry-toolbox for
nonlinear equations: the Camassa-Holm equation'', Program in Applied
Mathematics, University of Colorado at Boulder, preprint {\bf 160}}{(1993)}
\refjl{refFFb}{B. Fuchssteiner and A.S. Fokas}{Physica}{D4}{47}{1981}
\refjl{refFW}{B. Fornberg and G.B. Whitham}{Phil. Trans. R. Soc. Lond.
A}{289}{373--404}{1978}
\refjl{refG}{V.A. Galaktionov}{Diff. Int. Eqns.}{3}{863--874}{1990}
\refjl{refGDEKS}{V.A. Galaktionov, V.A. Dorodnytzin, G.G.
Elenin, S.P. Kurdjumov and A.A. Samarskii}{J. Sov. 
Math.}{41}{1222--1292}{1988}
\refjl{refGGKM}{C.S. Gardner, J.M. Greene, M.D. Kruskal and R. Miura}{Phys. 
Rev. Lett}{19}{1095--1097}{1967}
\refjl{refGP}{C. Gilson and A. Pickering}{\jpa}{28}{2871--2888}{1995}
\refeb{refH}{W. Hereman}{Euromath. Bull.}{1}{2}{45--79}{1994}
\refjl{refHS}{R. Hirota and J. Satsuma}{J. Phys. Soc.
Japan}{40}{611--612}{1976}
\refjl{refLW}{D. Levi and P. Winternitz}{J. Phys. A: Math.
Gen.}{22}{2915--2924}{1989}
\refjl{refLouii}{S.-Y. Lou}{Phys. Lett.}{A151}{133--135}{1990}
\refjl{refMak}{V.G. Makhankov}{Phys. Rep.}{35}{1--128}{1978}
\refbk{refM}{\liz}{``{\tt diffgrob2:} {\rm A symbolic algebra
package for analysing systems of PDE using Maple}'', {\tt ftp
ftp.ukc.ac.uk}}{login: anonymous, password: your
email address, directory: {\tt pub/maths/liz}}{1993}
\reftoap{refMC}{\liz\ and \pac}{{\rm ``Applications of the differential
algebra package {\tt diffgrob2} to classical symmetries of
differential equations'',}\ J. Symb. Comp.}{to appear}{1995}
\refjl{refMF}{\liz\ and E.D. Fackerell}{{\rm ``Differential
Gr\"obner Bases''}}{{\rm preprint} 92/108}{Macquarie University,
Sydney, Australia}{1992}
\refpp{refMB}{V. Marinakis and T.C. Bountis}{``On the integrability of a new
class of water wave equations'', preprint, Department of Mathematics,
University of Patras, Greece}{(1995)}
\refjl{refMcLO}{J.B. McLeod and P.J. Olver}{SIAM J. Math.
Anal.}{14}{488--506}{1983}
\refjl{refNC}{C. Nucci and \pac}{Phys. Lett.}{A164}{49--56}{1992}
\refjl{refO}{P.J. Olver}{{\sl ``Applications of Lie Groups to
Differential Equations''}, {\rm Second Edition} Graduate
Texts Math.}{107}{Springer, New York}{1993}
\refjl{refOlverb}{P.J. Olver}{Proc. R. Soc. Lond.
A}{444}{509--523}{1994}
\refpp{refOR}{P.J. Olver and P. Rosenau}{``Tri-Hamiltonian soliton-compacton
duality'', preprint, School of Mathematics, University of Minnesota,
Minneapolis}{(1995)}
\refjl{refPer}{H. Peregrine}{J. Fluid Mech.}{25}{321--330}{1966}
\refjl{refPucci}{E. Pucci}{\jpa}{25}{2631--2640}{1992}
\refjl{refRDG}{A. Ramani, B. Dorizzi and B. Grammaticos}{Phys. Rev.
Lett.}{49}{539--1541}{1982}
\refjl{refRRDG}{A.F. Ranada, A. Ramani, B. Dorizzi and B. Grammaticos}{J. Math.
Phys.}{26}{708--710}
{1985}\refjl{refR}{G.J. Reid}{J. Phys. A: Math. Gen.}{23}{L853--L859}{1990}
\refjl{refRi}{G.J. Reid}{Eur. J. Appl. Math.}{2}{293--318}{1991}
\refjl{refRose}{P. Rosenau}{Phys.
Rev. Lett.}{73}{1737--1741}{1993}
\refjl{refRH}{P. Rosenau and J.M. Hyman}{Phys.
Rev. Lett.}{70}{564--567}{1993}
\refbk{refS}{Yu.I. Shokin}{``The Method of Differential
Approximation''}{Springer, New York}{1983} 
\refjl{refV}{E.M. Vorob'ev}{Acta Appl. Math.}{24}{1--24}{1991}
\refjl{refWTC}{J. Weiss, M. Tabor and G. Carnevale}{J. Math.
Phys.}{24}{522--526}{1983}
\refjl{refWp}{G.B. Whitham}{Proc. R. Soc. Lond.
A}{299}{6--25}{1967}
\refbk{refWbk}{G.B. Whitham}{``Linear and Nonlinear Waves''}{Wiley,
New York}{1974}

\section{Introduction}
\setcounter{equation}{0}

In this paper we are concerned with symmetry reductions of the nonlinear
third order partial differential equation given by
\begin{equation}
 \Delta \equiv u_t -\epsilon u_{xxt} +2\kappa u_x-u u_{xxx}
-\alpha u u_x-\beta u_x u_{xx}=0, \label{fulleqn}
\end{equation}
where $\epsilon$, $\kappa$, $\alpha$ and $\beta$ are
arbitrary constants. Three special cases of (\ref{fulleqn})\
have appeared recently in the literature. Up to some rescalings, 
these are: (i), the Fornberg-Whitham equation \cite{\refFW,\refWp,\refWbk},
for the parameters $\epsilon=1$, $\alpha=-1$, $\beta=3$ and $\kappa=\tfr12$, 
(ii), the Rosenau-Hyman equation \cite{\refRH} for the parameters $\epsilon=0$,
$\alpha=1$, $\beta=3$ and $\kappa=0$, and
(iii), the Fuchssteiner-Fokas-Camassa-Holm equation
\cite{\refCH,\refCHH,\refFuch,\refFFb} for the parameters $\epsilon=1$,
$\alpha=-1$ and $\beta=2$.

The Fornberg-Whitham (FW) equation
\begin{equation}
u_t -u_{xxt} +u_x= u u_{xxx} -u u_x +3u_x u_{xx} \label{fw} 
\end{equation}
was used to look at qualitative behaviours of wave-breaking \cite{\refWp}.
It admits a wave of greatest height, as a peaked limiting
form of the travelling wave solution \cite{\refFW},
$$ u(x,t) = A \exp\left(-\tfr12 \vert x-\tfr43 t \vert\right),$$
where $A$ is an arbitrary constant.

The Rosenau-Hyman  (RH) equation
\begin{equation}
u_t =u u_{xxx}+u u_x +3u_x u_{xx}. \label{rh} 
\end{equation}
models the effect of nonlinear dispersion in the 
formation of patterns in liquid drops \cite{\refRH}.
It also has an unusual solitary wave solution, known as a ``compacton'',
$$ 
u(x,t) = \left \{
\begin{array}{cc}
 -\tfr83 c \cos^2\{\tfr14 (x-ct)\},& \mbox{if } \,|x-ct|\le 2\pi, \\[1.0ex]
            0,                     & \mbox{if } \,|x-ct|> 2\pi. 
\end{array} \right .
$$
These waves interact producing a ripple of
low amplitude compacton-anticompacton pairs.

The Fuchssteiner-Fokas-Camassa-Holm (FFCH) equation
\begin{equation}
 u_t -u_{xxt} +2\kappa u_x= u u_{xxx}
-3u u_x + 2u_x u_{xx}, \label{ch} 
\end{equation}
first arose in the work of Fuchssteiner and Fokas
\cite{\refFuch,\refFFb} using a bi-Hamiltonian approach;
we remark that it is only implicitly written in \cite{\refFFb} ---  see
equations (26e) and (30) in this paper --- though is explicitly written down in
\cite{\refFuch}. It has recently been rederived by Camassa and Holm
\cite{\refCH} from physical considerations as a model for dispersive shallow
water waves. In the case $\kappa=0$, it admits an unusual solitary wave solution
$$ u(x,t) = A\exp\left(-\vert x-ct \vert\right), $$
where $A$ and $c$ are arbitrary constants, which is called a ``peakon''. A
Lax-pair \cite{\refCH} and bi-Hamiltonian  structure \cite{\refFFb} have been
found for the FFCH equation (\ref{ch}) and so it appears to be completely integrable.
Recently the FFCH equation (\ref{ch}) has attracted considerable attention. In
addition to the aforementioned, other studies include 
\cite{\refCHH,\refCS,\refFoka,\refFokb,\refFokS,\refFuchb,\refGP,\refMB,\refOR}.

The FFCH equation (\ref{ch}) may be thought of as an integrable
modification of the regularized long wave (RLW) equation
\cite{\refBBM,\refPer}
\begin{equation}
u_{xxt} + u u_{x} - u_{t} - u_{x} = 0,\label{eqrlw}
\end{equation} sometimes known
as the Benjamin-Bona-Mahoney equation. However, in contrast to
(\ref{ch}), the RLW equation (\ref{eqrlw}) is thought \underbar{\it not} to be 
solvable by inverse scattering (cf., \cite{\refMcLO}); its solitary
wave solutions interact inelastically (cf., \cite{\refMak}) and only 
has finitely many local conservation laws \cite{\refOlverb}. However physically
it has more desirable properties than the celebrated Korteweg-de Vries
(KdV) equation
\begin{equation}
u_{t} + u_{xxx} + 6uu_x=0,\label{eqkdv}
\end{equation}
which was the first equation to be solved by inverse scattering
\cite{\refGGKM}. We remark that two other integrable variants of the
RLW equation (\ref{eqrlw}) are 
\begin{equation}
u_{xxt} + 2 u u_{t} -  u_{x}\partial_x^{-1}u_t -
u_{t} - u_{x} = 0,\label{eqswwa}
\end{equation}
where $\left(\partial_x^{-1} f\right)(x) = \int_x^\infty f(y)\,\mbox{d} y$,
which was introduced by Ablowitz, Kaup, Newell and Segur
\cite{\refAKNS}, and
\begin{equation}
u_{xxt} + u u_{t} -  u_{x}\partial_x^{-1}u_t -
u_{t} - u_{x} = 0,\label{eqswwb}
\end{equation}
which was discussed by Hirota and Satsuma \cite{\refHS}.
We also note that (\ref{ch}), with $\kappa=\tfr12$, (\ref{eqrlw}), (\ref{eqswwa})
and (\ref{eqswwb}) all have the same linear
dispersion relation $\omega(k)=-k/(1+k^2)$ for the complex exponential
$u(x,t)\sim\exp\{\mbox{i}[kx+\omega(k)t]\}$.

Recently, Gilson and Pickering \cite{\refGP} have shown that no equation in the
entire class of equations (\ref{fulleqn})\ will  satisfy the necessary conditions of
either the  Painlev\'e PDE test due to Weiss, Tabor and Carnevale \cite{\refWTC}
or the Painlev\'e ODE test due to Ablowitz, Ramani and Segur
\cite{\refARS,\refARSi} to be solvable by inverse scattering. However, the
integrable FFCH equation (\ref{ch}) does possess the ``weak Painlev\'e'' property
(cf., \cite{\refRDG,\refRRDG}), as does the FW equation (\ref{fw}).

All these special travelling wave solutions 
are essentially exponential solutions, or sums of exponential
solutions, and thus would suggest some sort of linearity
in the differential equation. This is discussed by
Gilson and Pickering \cite{\refGP}, who show that (\ref{fulleqn}), with
$\alpha\not=0$ and $\beta(1+\beta)\not=0$, can be written as
\begin{equation}
 \left( \beta u_x +u {\partial_x} +\epsilon
{\partial_t}\right) ( u_{xx} -\mu^2 u -2\kappa/\beta) =0,
\label{ufactor} 
\end{equation} 
where $\partial_x\equiv{\partial/\partial x}$,
$\partial_t\equiv{\partial/\partial t}$ and $\mu^2=-\alpha/(1+\beta)$, provided
that $\epsilon\alpha+\beta +1=0$, which includes the FFCH equation (\ref{ch}). For
the travelling wave reduction,
$$u=w(z),\qquad z=x-ct,$$ the resulting ordinary differential
equation is
\begin{equation}
 (2\kappa -c) w'+\epsilon c w'''-ww'''-\alpha ww'- \beta w'w''=0, \label{travwav}
\end{equation} 
where $'\equiv{\mbox{d}/ \mbox{d} z}$, which also may be factorised as
\begin{equation}
\left [ \beta w'+(w-\epsilon c ){ \mbox{d} \over  \mbox{d} z} \right ] 
(w''-\mu^2 w+\gamma)=0, \label{wfactor}
\end{equation}
provided that
$$\mu^2=-{\alpha\over 1+\beta},\qquad \beta(1+\beta)\gamma -2\kappa(1+\beta) +
c(1+\beta+\alpha\epsilon)=0.$$ This includes all three special cases
(\ref{fw})--(\ref{ch}); since $\beta(1+\beta)$ is strictly non-zero in these three cases
then a suitable $\gamma$ can always be found.

Furthermore, if $1+\beta+\alpha\epsilon=0$ and $\epsilon\not=0$, then (\ref{fulleqn})
with $\kappa=0$ possesses the ``peakon'' solution
$$ u(x,t) = A \exp\left(-\epsilon^{-1/2}\vert x-ct \vert\right),$$ where $A$ and
$c$ are arbitrary constants.  More generally, if $\alpha/(1+\beta)<0$,
$1+\beta+\alpha\epsilon\not=0$ and
$\kappa\not=0$, then (\ref{fulleqn})\ possesses the solution
$$ u(x,t) = A \exp\left\{-\left(-\,{\alpha\over1+\beta}\right)^{1/2}\vert x-ct
\vert\right\},\qquad c={2(1+\beta)\kappa\over 1+\beta+\alpha\epsilon},$$ where
$A$ is an arbitrary constant. If $\alpha/(1+\beta)>0$, $\beta\not=-1$ and
$\alpha\beta\not=0$, then (\ref{fulleqn})\ possesses the ``compacton'' solution
$$ u(x,t) = {2[2(1+\beta)\kappa-(1+\beta+\alpha\epsilon)c]\over\alpha\beta}
\cos^2\left\{\tfr12\left(\alpha\over1+\beta\right)^{1/2}(x-ct)\right\},$$ where
$c$ is an arbitrary constant.

The classical method for finding symmetry reductions of partial differential
equations is the Lie group method of infinitesimal transformations. As this
method is entirely algorithmic, though often both tedious and virtually
unmanageable manually, symbolic manipulation programs have been developed to aid
the calculations. An excellent survey of the different packages available and a
description of their strengths and applications is  given by Hereman
\cite{\refH} (see also his contribution in this volume). 
In this paper we use the  {\rm MACSYMA} package {\tt symmgrp.max}
\cite{\refCHW} to calculate the determining equations.

In recent years the nonclassical method due to Bluman and Cole \cite{\refBC} (in
the sequel referred to as the ``nonclassical method''), sometimes referred to as
the ``method of partial symmetries of the first type'' \cite{\refV}, or the
``method of  conditional symmetries'' \cite{\refLW}, and the direct method 
due to Clarkson and Kruskal \cite{\refCK} have been used to generate many new
symmetry reductions and exact solutions for several physically significant
\pdes\ that are not obtainable using the classical Lie method (cf.,
\cite{\refPAC} and the references therein). The nonclassical method is a
generalization of the classical Lie method, whereas the direct method is an
ansatz-based approach  which involves no group theoretic techniques. Nucci and
Clarkson \cite{\refNC} showed that for the Fitzhugh-Nagumo equation the 
nonclassical method is more general than the direct method, since they
demonstrated the existence of a solution of the Fitzhugh-Nagumo equation,
obtainable using the nonclassical method but not using the direct method. 
Subsequently Olver [\refOlverb] (see also [\refABH,\refPucci]) has proved  the
general result  that for a scalar equation, every reduction obtainable using the
direct method is also obtainable using the nonclassical method. Consequently we
use the nonclassical method in this paper rather than the direct method.

Symmetry reductions and exact solutions have several  different important
applications in the context of differential equations. Since solutions of
\pdes\  asymptotically tend to solutions of lower-dimensional equations obtained
by symmetry reduction, some of these special solutions will illustrate important
physical phenomena. In particular, exact solutions arising from symmetry methods
can often be used effectively to study properties such as asymptotics and
``blow-up'' (cf., \cite{\refG,\refGDEKS}). Furthermore, explicit solutions (such
as those found by symmetry methods)  can play an important role in the design
and testing of numerical integrators; these solutions provide an important
practical check on the accuracy and reliability of 
such integrators (cf., \cite{\refA,\refS}). 

Classical symmetries of differential equations are found in practice by a
two-step process.  The first involves finding the determining equations for the
infinitesimals of the group action.  These determining equations form an
overdetermined, linear system of {\pdes}.  The second step involves integrating
this system. The first step is entirely algorithmic, and has been implemented
in all the commercial symbolic manipulation languages (cf., \cite{\refH}).
The second step involves heuristic integration procedures which have been
implemented in some symbolic manipulation programs and are largely successful,
though not infallible.  Commonly, the overdetermined systems to be solved are
simple, and heuristic integration is both fast and effective.  However, there
are three areas where heuristics can break down (cf., \cite{\refMC} for further
details and examples).
\begin{itemize}
\item[1.]{\it Arbitrary parameters and functions}. If the \pde\ whose
symmetries are sought involves arbitrary parameters, such as (1.1) or more
generally, arbitrary functions,  heuristics yield usually the general solution,
and miss those special cases of the parameters and arbitrary functions where
additional symmetries exist.
\item[2.]{\it Termination}. Heuristic algorithms are not guaranteed to
terminate, and may become trapped in infinite loops for some examples.
\item[3.]{\it Too difficult to solve}.  The system may not be solvable
by the heuristic.  The heuristic will then attempt to represent the general
solution in terms of functions satisfying certain conditions, but may
give up before a useful representation is obtained.
\end{itemize}
These problems are addressed by use differential Gr\"obner bases
(DGBs) which we describe below.

The method used to find solutions of the determining equations in the
nonclassical method is that of DGBs, defined to
be a basis \ss\ of the differential ideal generated by the system such that
every member of the ideal pseudo-reduces to zero with respect to \ss. This method
provides a systematic framework for finding integrability and compatibility
conditions of an overdetermined system of \pdes. It avoids the problems of
infinite loops in reduction processes and yields, as far as is currently
possible, a ``triangulation'' of the system from which the solution set can
be derived more easily \cite{\refCMi,\refMF,\refR,\refRi}. In a sense,
a DGB provides the maximum amount of information possible using elementary
differential and algebraic processes in finite time.

In pseudo-reduction, one must, if necessary, multiply the expression being
reduced by differential (non-constant) coefficients of the highest
derivative terms of the reducing equation, so that the algorithms used
will terminate \cite{\refMF}. In practice, such coefficients are assumed to be
non-zero, and one needs to deal with the possibility of them being zero
separately. These are called singular cases.

The triangulations of the systems of \deqs\ for \inls\ arising in the
nonclassical method in this paper were all performed using the {\rm MAPLE}
package {\tt diffgrob2} \cite{\refM}. This package was written specifically
to handle nonlinear equations of polynomial type. All calculations are strictly
`polynomial', that is, there is no  division. Implemented there are the
Kolchin-Ritt algorithm using pseudo-reduction instead of reduction, and extra
algorithms needed to calculate a DGB (as far as possible using the current
theory), for those cases where the Kolchin-Ritt algorithm is not sufficient
\cite{\refMF}. The package was designed to be used interactively as well as
algorithmically, and much use is made of this fact here. It has proved useful
for solving many fully nonlinear systems \cite{\refCMi--\refCMiv}. 

In the following sections we shall consider the cases $\epsilon=0$ and
$\epsilon\not=0$, when we set $\epsilon=1$ without loss of generality,
separately because the presence or lack of the corresponding third order term 
is significant. In \S 2 we find the classical Lie group of symmetries and
associated reductions of (\ref{fulleqn}). In \S3 we discuss the nonclassical
symmetries and reductions of (\ref{fulleqn})\ in the generic case. In \S4 we
consider special cases of the the nonclassical method in the so-called
$\tau=0$; in full generality this case generates a single equation which
is considerably more complex than our original equation! In
\S5 we discuss our results.

\section{Classical symmetries}
\setcounter{equation}{0}

To apply the classical method we consider
the one-parameter Lie group of infinitesimal 
transformations in ($x,t,u$) given by
\begin{equation}
\label{trans}
\begin{array}{ccl}
x^*&=&x+\varepsilon \xi (x,t,u)  + O (\varepsilon^2),  \\[1.0ex]
t^*&=&t+\varepsilon \tau (x,t,u) + O (\varepsilon^2),  \\[1.0ex]
u^*&=&u+\varepsilon \phi (x,t,u) + O (\varepsilon^2),  
\end{array}
\end{equation}
where $\varepsilon$ is the group parameter. Then one requires that
this transformation leaves invariant the set
\begin{equation}
S_{\Delta} \equiv \{ u(x,t) : \Delta =0 \} \label{sdel}
\end{equation}
of solutions of (\ref{fulleqn}). This
yields an overdetermined, linear system of equations for the
infinitesimals $\xi (x,t,u),\tau (x,t,u),\phi (x,t,u)$. The
associated Lie algebra is realised by vector fields of the form
\begin{equation}
 {\bf v} = \xi (x,t,u) {\partial_x} +\tau (x,t,u) {\partial_t} +
 \phi (x,t,u) {\partial_u}. \label{vecfi}
\end{equation}
Having determined the infinitesimals, the symmetry variables are
found by solving the characteristic equation
\begin{equation}
 {\mbox{d} x \over \xi (x,t,u)} = {\mbox{d} t \over \tau (x,t,u)}=
 {\mbox{d} u \over \phi (x,t,u)}, \label{chareq}
\end{equation}
which is equivalent to solving the \isc\
\begin{equation}
 \psi \equiv \xi (x,t,u) u_x+\tau (x,t,u) u_t -
 \phi (x,t,u) =0. \label{invsc}
\end{equation}
The set $S_{\Delta}$ is invariant under the transformation (\ref{trans}) provided
that $ {\rm pr}^{(3)} {\bf v} (\Delta) |_{\Delta \equiv 0} =0$
where ${\rm pr}^{(3)}{\bf v}$ is the third prolongation of the vector field
(\ref{vecfi}), which is given explicitly in terms of $\xi,\tau$ and $\phi$ (cf.\
\cite{\refO}). This procedure  yields the determining equations. There are two
cases to consider.

\subsection{$\epsilon=0$}

In this case using the {\rm MACSYMA} package {\tt symmgrp.max}
we obtain the following system of ten determining equations 
\begin{equation}
\label{clez}
\begin{array}{@{\hspace{0.0ex}}l@{\hspace{0.0ex}}}
\tau_{u}=0, \quad 
\tau_{x}=0, \quad
\xi_{u} =0, \quad
u\phi_{uuu}+\beta\phi_{uu}=0, \quad
3u^2\phi_{uu}+\beta u\phi_{u}-\beta\phi=0,\\[1.0ex]  
3u\phi_{xu}-3u\xi_{xx}+\beta\phi_{x}=0, \quad 
3u\phi_{xuu}+2\beta\phi_{xu}-\beta\xi_{xx}=0,\\[1.0ex]
\tau_{t}u-3\xi_{x}u+\phi=0,\quad
\phi_{xxx}u+(\alpha u-2\kappa)\phi_{x}-\phi_{t}=0,\\[1.0ex] 
3u^2\phi_{xxu}+\beta u\phi_{xx}+2\kappa\phi -u^2\xi_{xxx}
      +(2\alpha u^2-4\kappa u)\xi_{x}+u\xi_{t}=0.
\end{array}
\end{equation}
Next applying the {\tt reduceall} algorithm in the 
{\rm MAPLE} package {\tt diffgrob2} to this system yields
$$
\begin{array}{l}
(2+\beta)\xi_{xx}=0,\quad  
(2+\beta)[\alpha u\xi_{xt}+\xi_{tt} -2\kappa\xi_{xt}]=0, \\[1.0ex]
\xi_{u}=0, \quad \tau_{x}=0, \quad  \tau_{u}=0,\\[1.0ex]    
2\alpha u\xi_{x}+2\kappa\xi_{x}+\xi_{t} -2\kappa\tau_{t}=0,\quad 
(2+\beta)[2\kappa\phi+(2\alpha u^2-4\kappa u)\xi_x+u\xi_t]=0.
\end{array}
$$
This is simple enough to solve; there is no need to do the full Kochin-Ritt
algorithm in this case. The output shows that there are three special values of
the parameters, namely
$\alpha=0$, $\beta=-2$ and $\kappa=0$, and combinations thereof. It transpires
that the special case $\beta=-2$ is purely an artefact.
For the three special cases (a) $\alpha=0$, $\kappa\not=0$, (b)
$\alpha\not=0$, $\kappa=0$ and (c) $\alpha=\kappa=0$, applying the {\tt
reduceall} algorithm of {\tt diffgrob2} to (\ref{clez}) yields\\

\noindent
\begin{tabular}{cl@{\hspace{5.0ex}}l}
(a)&$\alpha=0$,\quad $\kappa\not=0$ 
   &$\xi_{xx}=0,\quad \xi_{tt} -2\kappa\xi_{xt}=0,\quad\xi_{u}=0,$\\[1.0ex]
  &&$\tau_{x}=0,\quad 2\kappa\xi_{x}+\xi_{t}-2\kappa\tau_{t}=0, \quad\tau_{u}=0,$\\[1.0ex]
  &&$2\kappa\phi+(2\alpha u^2-4\kappa u)\xi_x+u\xi_t=0.$\\[2.0ex]
\end{tabular}
\begin{tabular}{cl@{\hspace{5.0ex}}l}
(b)&$\alpha\not=0$,\quad $\kappa=0$ 
   &$2\alpha u\xi_{x} +\xi_{t}=0,\quad \xi_{tt}=0,\quad\xi_{u}=0,$\\[1.0ex]
  &&$\tau_{x}=0,\quad\tau_{tt}=0,\quad \tau_{u}=0,$\\[1.0ex]
  &&$2\kappa\phi+(2\alpha u^2-4\kappa u)\xi_x+u\xi_t=0.$\\[2.0ex]
\end{tabular}\\
\begin{tabular}{cl@{\hspace{5.0ex}}l}
(c)&$\alpha=\kappa=0$ 
   &$ \xi_{xx}=0,\quad\xi_{t}=0,\quad\xi_{u}=0,$\\[1.0ex]
  &&$\tau_{x}=0,\quad\tau_{tt}=0,\quad\tau_{u}=0,$\\[1.0ex]
  &&$\phi-3u\xi_x+u\tau_t=0$.\\[1.0ex]
\end{tabular}\\

Hence we obtain the following infinitesimals:\\

\noindent
{\sc Case} 2.1(i) $\alpha\not=0$ and
$\kappa\not=0$
\begin{equation}
\xi= 2\kappa c_3 t + c_1,\qquad\tau=
c_3 t+ c_2,\qquad\phi=-c_3 u.\label{kzii}
\end{equation}  
\noindent
{\sc Case} 2.1(ii)  $\alpha=0$ and $\kappa\not=0$
\begin{equation}
\xi= c_3x+2\kappa(c_4-c_3) t + c_1,\qquad\tau=c_4 t+
c_2,\qquad\phi=(3c_3-c_4) u.\label{kziii}
\end{equation}   
\noindent
{\sc Case} 2.1(iii) $\alpha\not=0$ and $\kappa=0$
\begin{equation}
\xi= c_1,\qquad\tau=c_3 t+ c_2,\qquad\phi=-c_3 u.\label{knii}
\end{equation}
\noindent
{\sc Case} 2.1(iv)  $\alpha=0$ and $\kappa=0$
\begin{equation}
\xi= c_3x+c_1,\qquad\tau=c_4 t+c_2,\qquad\phi=(3c_3-c_4)
u,\label{kniii}
\end{equation}  where $c_1,c_2,c_3,c_4$ are arbitrary constants. 

Solving the invariant surface condition (\ref{invsc})\ yields four different 
canonical reductions:\\

\noindent{\bf Reduction 2.1}
$\alpha$ and $\kappa$ arbitrary. 
If $c_3=c_4=0$ in (\ref{kzii})--(\ref{kniii}) we may set $c_1=c$ and
$c_2=1$ and we obtain the travelling wave reduction
$$ u(x,t)= w(z),\qquad z=x-c t,$$ where $w(z)$ satisfies
$$ w w''' +\alpha w w' +\beta w'w''+ (c-2\kappa)w'=0. $$ This can be
integrated to yield
$$ w w''+\tfr12 (\beta-1) (w')^2 +\tfr12 \alpha w^2 + (c-2\kappa)w=A,  $$
where $A$ is an \ac. Multiplying this by $w^{\beta-2} w'$ and integrating again
yields
\begin{equation}
 (w')^2 +{\alpha\over1+\beta} w^2 + {2(2\kappa -c)\over\beta}
w={2A\over\beta-1}+B w^{1-\beta},\label{eqeztw}
\end{equation}
where $B$ is an \ac, for
$\beta\not=-1,0,1$.  Generally if $\beta\not=-1,0,1$, then (\ref{eqeztw}) is solvable
using quadratures, though for certain special values of the parameters there are
explicit solutions. For example (i), if $\beta=-2$ or
$\beta=-3$, then (\ref{eqeztw}) is solvable in terms of Weierstrass or Jacobi
elliptic functions, respectively, (ii), if $B=0$, then (\ref{eqeztw}) is solvable in
term of trigonometric functions, and (iii), if
$c=2\kappa$ and $\beta=3$, then $w(z)$ can be expressed in terms of
trigonometric functions via the transformation $w(z)=v^{1/2}$.

In the special cases $\beta=-1,0,1$ we obtain the equations
$$
\begin{array}{l}
(w')^2 +\alpha w^2\ln w+2(2\kappa -c) w= B w^2 - A, \\[1.0ex]
(w')^2 +\alpha w^2+2(2\kappa -c) w\ln w= B w - 2A, \\[1.0ex]
(w')^2 + \alpha w^2 +2(2\kappa -c) w= B- A \ln w,
\end{array}
$$
respectively, with $A$ and $B$ arbitrary functions. If the coefficient of $\ln
w$ in these equations is zero, then $w(z)$ is expressible in terms of elementary
functions, otherwise in terms of quadratures.\\

\noindent{\bf Reduction 2.2} $\alpha\not=0$, $\kappa$ arbitrary.
If $c_3\not=0$ in (\ref{kzii}) and (\ref{knii}) we may set $c_3=1$,
$c_1=c$ and $c_2=0$, without loss of generality, and obtain 
the reduction
\begin{equation}
u(x,t) = w(z) t^{-1},\qquad z= x  -c\ln t -2\kappa
t,\label{redIIiiw}
\end{equation}
where $w(z)$ satisfies
$$ w w'''+\beta w' w''+\alpha w w'+c w' + w =0.$$ 
Also if $c_3=0$ and $c_4\not=0$ in (\ref{kziii}) we may set 
$c_4=1$, $c_1=c$ and $c_2=0$, without loss of generality,
and obtain the same reduction (\ref{redIIiiw}).\\

\noindent{\bf Reduction 2.3} $\alpha=0$, $\kappa$ arbitrary. 
If $c_3\not=0$ and $c_4\not=0$ in (\ref{kziii}) and (\ref{kniii}), we may set
$c_3=m+\tfr13$,
$c_4=1$ and $c_1=c_2=0$, without loss of generality, and obtain 
the reduction
$$ u(x,t) = w(z) t^{3m},\qquad z=(x-2\kappa t)t^{-m-1/3},$$ where $w(z)$
satisfies
$$ w w''' +\beta w'w''+(m+\tfr13) zw'-3m w=0.$$ 
\vspace{1.0ex}

\noindent{\bf Reduction 2.4} $\alpha=0$, $\kappa$ arbitrary. 
If $c_3\not=0$ and $c_4=0$ in (\ref{kziii}) and (\ref{kniii}), we may set $c_3=m$,
$c_1=2\kappa$ and $c_2=1$, without loss of generality, and obtain 
the reduction
$$ u(x,t) = w(z) \mbox{e}^{3m t}, \qquad
 z= (x-2\kappa t) \mbox{e}^{-m t},
$$ 
where $w(z)$ satisfies
$$ w w''' +\beta w' w'' +m z w' -3m w =0.$$

\subsection{$\epsilon=1$}
In this case we obtain the following system of eleven determining equations:
\begin{equation}
\label{clen}
\begin{array}{l}
\tau_{u}=0, \quad
\tau_{x}=0, \quad
\xi_u=0, \quad
\phi_{uu}=0,\quad
2 \phi_{xu} - \xi_{xx}=0, \\[1.0ex]
\beta(u\phi_{u}  - \phi + \xi_{t})=0, \quad 
\phi + u\tau_{t} -u \xi_{x}  - \xi_{t}=0,\\[1.0ex]
3u \phi_{xu} + \phi_{tu}+ \beta \phi_{x} - 3u \xi_{xx}- 2 \xi_{xt} =0, \\[1.0ex]
u\phi_{xxu} + \phi + u\tau_{t}  - 3 \xi_{x} u  - \xi_{t} =0, \\[1.0ex]
u\phi_{xxx}  + \phi_{xxt} -\phi_{t} +(\alpha u -2\kappa) \phi_{x}=0, \\[1.0ex]
3  u^{2}\phi_{xxu} + 2 u\phi_{xtu}  +\beta u\phi_{xx} 
  + 2 \kappa \phi
  - u^{2} \xi_{xxx} 
 - u\xi_{xxt} \\[1.0ex]
{~}\hfill 
 +(2\alpha u^2-4\kappa)\xi_{x} 
+[(\alpha+1) u-2\kappa]\xi_{t} =0.
\end{array}
\end{equation}
As in the previous case, we apply the {\tt reduceall} algorithm in the {\rm
MAPLE} package {\tt diffgrob2}, to this system, which yields
$$
\begin{array}{l}
\xi_x=0,\quad (\alpha+1)\xi_{tt}=0,\quad \xi_u=0,\\[1.0ex]
\tau_x=0,\quad (\alpha+1)\tau_{tt}=0,\quad \tau_u=0,\\[1.0ex]
2\kappa\phi = [2\kappa - (\alpha+1) u ]\xi_t.
\end{array}
$$ 
This shows that there are two special
values of the parameters, namely $\alpha=-1$ and $\kappa=0$.  For the three
special cases (a) $\alpha=-1$, $\kappa\not=0$, (b)
$\alpha\not=-1$, $\kappa=0$ and (c) $\alpha=-1$, $\kappa=0$, applying the {\tt
reduceall} algorithm of {\tt diffgrob2} to (\ref{clen}) yields\\

\begin{tabular}{cl@{\hspace{5.0ex}}l}
(a)& $\alpha=-1$,\quad $\kappa\not=0$  
   & $\xi_x=0,\quad \xi_{tt}=0,\quad \xi_u=0,$\\[1.0ex]
  && $\tau_x=0,\quad \tau_{t}=0,\quad \tau_u=0,$\\[1.0ex]
  && $\phi = \xi_t.$\\[2.0ex]
(b)& $\alpha\not=-1$,\quad $\kappa=0$ 
   & $\xi_x=0,\quad \xi_{t}=0,\quad \xi_u=0,$\\[1.0ex]
  && $\tau_x=0,\quad \tau_{tt}=0,\quad \tau_u=0,$\\[1.0ex]
  && $\phi = -u\tau_t.$\\[2.0ex]
(c)& $\alpha=-1$, $\kappa=0$ 
   & $\xi_{x}=0,\quad\xi_{tt}=0,\quad\xi_{u}=0,$\\[1.0ex]
  && $\tau_{x}=0,\quad\tau_{tt}=0,\quad\tau_{u}=0,$\\[1.0ex] 
  && $\phi-u\xi_t+u\tau_t=0.$\\[1.0ex]
\end{tabular}\\

Hence we obtain the following infinitesimals:\\

\noindent {\sc Case} 2.2(i) $\alpha\not=-1$,
$\kappa\not=0$
\begin{equation}
\xi=c_3 t + c_1,\quad \tau=
 {(1+\alpha)c_3 t\over2\kappa} + c_2,\quad
\phi=c_3\left[1-{(1+\alpha)u\over2\kappa}\right].\label{kzi}
\end{equation}

\noindent {\sc Case} 2.2(ii) $\alpha=-1$, $\kappa\not=0$
\begin{equation}
\xi=c_3 t + c_1,\quad \tau=c_2,\quad
\phi=c_3.\label{Kzii}\end{equation}

\noindent {\sc Case} 2.2(iii) $\alpha\not=-1$, $\kappa=0$
\begin{equation}
\xi=c_1,\quad \tau=c_3 t + c_2,\quad \phi=-c_3u.
\label{Kziii}\end{equation}

\noindent {\sc Case} 2.2(iv) $\alpha=-1$, $\kappa=0$
\begin{equation}
\xi=c_3 t + c_1,\quad \tau=c_4 t + c_2,\quad
\phi=c_3-c_4u,\label{kziv}\end{equation}
where $c_1,c_2,c_3,c_4$ are arbitrary constants.\\

There are four canonical reductions.\\

\noindent{\bf Reduction 2.5} $\alpha$ and $\kappa$ arbitrary.
If in (\ref{kzi})--(\ref{kziv}) $c_3=c_4=0$, we may set $c_1=c$ and
$c_2=1$ without loss of generality. Thus we obtain the reduction
$$ u(x,t)=w(z)+c, \quad z=x-c t, $$ where $w(z)$ satisfies
$$ w w'''+\beta w'w''+\alpha ww'=[2\kappa-(1+\alpha)c]w'. $$ This can be
integrated to yield
$$ w w''+\tfr12(\beta+1)(w')^2+\tfr12\alpha w^2=[2\kappa-(1+\alpha)c]w + A,
$$ where $A$ is an \ac. Then multiplying through by $w^{\beta-2} w'$ and
integrating again yields
\begin{equation}
 (w')^2+{2\alpha w^2\over \beta+1} ={2[2\kappa-(1+\alpha)c]w\over\beta} +
{2A\over \beta-1} + Bw^{1-\beta},\label{eqentw}
\end{equation} 
provided that $\beta\not=-1,0,-1$.  Generally if $\beta\not=-1,0,1$, then
(\ref{eqentw}) is solvable using quadratures, though for certain special values of
the parameters, there are explicit solutions. For example (i), if $\beta=-2$ or
$\beta=-3$, then (\ref{eqentw}) is solvable in terms of Weierstrass or Jacobi
elliptic functions, respectively, (ii) if $B=0$, then (\ref{eqentw}) is solvable in
term of trigonometric functions, and (iii) if
$(1+\alpha)c=2\kappa$ and $\beta=3$, then $w(z)$ can be expressed in terms of
trigonometric functions via the transformation $w(z)=v^{1/2}$.

In the special cases $\beta=-1,0,1$ we obtain the following equations,
$$
\begin{array}{l}
(w')^2+2\alpha w^2\ln w= -2[2 \kappa -(1+\alpha)c]w-A+B w^2,\\[1.0ex]
(w')^2+2\alpha w^2 = -2[2 \kappa -(1+\alpha)c]w\ln w-2A+B w^2,\\[1.0ex]
(w')^2+2\alpha w^2 = -2[2 \kappa -(1+\alpha)c]w+2A\ln w+B w^2,
\end{array}
$$ 
respectively,
where $A$ and $B$ are arbitrary constants. If the coefficient of $\ln w$ in
these equations is zero, then $w(z)$ is expressible in terms of elementary
functions, otherwise in terms of quadratures.\\

\noindent{\bf Reduction 2.6} $\alpha\not=-1$, $\kappa$ arbitrary. 
If $c_3\not=0$ in (\ref{kzi}), we may set
$c_3=1$, $c_2=0$ and $c_1=2\kappa c/(1+\alpha)$, without loss
of generality. Thus we obtain the reduction
\begin{equation}
 u(x,t) = {w(z)+c\over t} +{2\kappa\over1+\alpha},\quad z=x-{2\kappa
t\over1+\alpha} -c \ln t, \label{redIIIiiw}
\end{equation}
where $w(z)$ satisfies
\begin{equation}
 ww'''+\beta w'w''- w''+\alpha ww'+(\alpha +1)c w'+w+c=0.
\label{redIIIiieq}
\end{equation}
If $c_3\not=0$ in (\ref{Kziii}) we may set
$c_3=1$, $c_1=c$ and $c_2=0$ to obtain the reduction
(\ref{redIIIiiw}) with $\kappa=0$.\\

\noindent{\bf Reduction 2.7} $\alpha=-1$, $\kappa\not=0$. 
If $c_3\not=0$ in (\ref{Kzii}) then we
set $c_3=m$, $c_1=0$ and $c_2=1$, without loss of generality. Thus
we obtain the reduction
\begin{equation}
 u(x,t)=w(z) +m t,\quad z=x-\tfr12m t^2, \label{redIIIiiiw}
\end{equation}
where $w(z)$
satisfies
$$ ww'''+\beta w'w''- ww'-2\kappa w'-m=0,$$  which may be integrated to
yield
\begin{equation}
 w w'' +\tfr12 (\beta-1) (w')^2 -\tfr12 w^2 -2\kappa w -m
z=A,\label{redIIIiiieq}
\end{equation}
where $A$ is an \ac.\\

\noindent{\bf Reduction 2.8} $\alpha=-1$, $\kappa=0$. 
If $c_3\not=0$ and $c_4\not=0$ in (\ref{kziv}) we may set
$c_3=m$, $c_4=1$, $c_1=c$ and $c_2=0$, without loss of
generality. Thus we obtain the reduction 
\begin{equation}
u(x,t)={w(z)+c\over t} +m,\quad z=x-m t - c\ln t,  \label{redIIIivw}
\end{equation} 
where $w(z)$ satisfies
\begin{equation}
 ww''' +\beta  w'w''-w''- ww' +w+c =0.\label{redIIIiveq}
\end{equation}

\section{Nonclassical symmetries ($\tau\not=0$)}
\setcounter{equation}{0}

In the nonclassical method one requires only the 
subset of $S_{\Delta}$ given by
\begin{equation}
S_{\Delta,\psi} = \{ u(x,t) : \Delta (u) =0, 
\psi (u) =0 \}, \label{sdelpsi} 
\end{equation}
where $S_{\Delta}$ is defined in (\ref{sdel})\ and $\psi=0$ is
the \isc\ (\ref{invsc}), to be invariant under the transformation
(\ref{trans}). The usual method of applying the nonclassical method
(e.g. as described in \cite{\refLW}), involves applying the 
prolongation ${\rm pr}^{(3)} {\bf v}$ to the system 
composed of (\ref{fulleqn})\ and the \isc\ (\ref{invsc})\ and requiring
that the resulting expressions vanish for $u\in S_{\Delta,\psi}$,
i.e.
\begin{equation}
 {\rm pr}^{(3)} {\bf v} (\Delta) |_{\Delta=0,\psi=0}=0,
\quad {\rm pr}^{(1)} {\bf v} (\psi) \vert_{\Delta=0,\psi=0}=0.
\label{twoprol} 
\end{equation} 
It can well known that the latter vanishes identically when $\psi=0$
without imposing any conditions upon $\xi$, $\tau$ and $\phi$. To apply the
method in practice we advocate the algorithm described in \cite{\refCMiii} for
calculating the determining equations, which avoids difficulties arising from
using differential consequences of the \isc\ (\ref{invsc}). 

In the canonical case when $\tau\not=0$ we set $\tau=1$
without loss of generality. We proceed by eliminating
$u_t$ and $u_{xxt}$ in (\ref{fulleqn})\ using the \isc\ (\ref{invsc})\
which yields
\begin{equation}
\label{delstar}
\begin{array}{l}
  \epsilon\xi u_{xxx} -uu_{xxx} +3\epsilon\xi_u u_x u_{xx}
               -\beta u_x u_{xx}-\epsilon\phi_u u_{xx} 
               +2\epsilon\xi_x u_{xx} \\[0.7ex]
               +\epsilon\xi_{uu} u_x^3 
  -\epsilon\phi_{uu} u_x^2  
 +2\epsilon\xi_{xu} u_x^2-\alpha u u_x -2\epsilon\phi_{xu} u_x 
 +2 \kappa u_x \\[0.7ex]
 +\epsilon\xi_{xx} u_x 
 -\epsilon\phi_{xx}
 +\phi-\xi u_x =0. 
\end{array}
\end{equation}
We note that this equation now involves the infinitesimals $\xi$ and $\phi$
that are to be determined. Then we apply the classical Lie
algorithm to (\ref{delstar}) using the third
prolongation $ {\rm pr}^{(3)} {\bf v}$ and eliminating
$u_{xxx}$ using (\ref{delstar}). It should be noted that the 
coefficient of $u_{xxx}$ is ($\xi-\epsilon u$). Therefore, if this is zero
the removal of $u_{xxx}$ using (\ref{delstar}) is invalid 
and so the next highest derivative term, $u_{xx}$, should
be used instead. We note again that this has a coefficient, $\beta-3$,
and so that in the case $\xi=u$ one needs to calculate
the \deqs\ for the cases $\beta\ne 3$ and $\beta=3$ separately. Continuing in
this fashion, there is a cascade of cases to be considered.  In the remainder
of this section, we consider these cases in turn.  First, however, we discuss
the case given by $\epsilon=0$.

\subsection{$\epsilon=0$} 

The first determining equation gives $\xi_u=0$,
and substituting this into the other seven determining equations yields
\begin{equation}
\label{nez}
\begin{array}{l}
\phi_{uuu} u + \beta \phi_{uu}=0, \quad
3 \phi_{xuu} u + 2 \beta \phi_{xu} - \beta \xi_{xx}=0,\\[2.0ex]
3 \phi_{uu} u^{2} + \beta \phi_{u} u - \beta \phi=0,\quad
3 \phi_{xu} u - 3 \xi_{xx} u + \beta \phi_{x}=0,\\[2.0ex]
\phi_{t} u-\phi_{xxx}u^{2}-\alpha\phi_{x}u^{2}
 + 2 \kappa \phi_{x} u + 3 \xi_{x} \phi u - \phi^{2}=0,\\[2.0ex]
3 \phi_{xxu} u^{2} - \xi_{xxx} u^{2} + 2 \alpha \xi_{x} u^{2}
  + \beta \phi_{xx} u - 4 \xi_{x} \kappa u
   + 3 \xi \xi_{x} u \\[1.0ex]
 + \xi_{t} u + 2 \kappa \phi - \xi \phi =0.
\end{array}
\end{equation}
It is quite straightforward to solve these equations and so we obtain the
following \inls: 
(a), if $\alpha\not=0$
$$ 
\begin{array}{l}
{\hbox to 30pt{(i)\hfill}} \displaystyle \quad \xi=2\kappa +{c_1\over t+c_2}, \quad
 \phi={- u \over t+c_2},\\[3.0ex]
{\hbox to 30pt{(ii)\hfill}} \displaystyle \quad \xi=c_1, \quad \phi=0, 
\end{array}
$$ 
and (b), if $\alpha=0$
$$ 
\begin{array}{l}
{\hbox to 30pt{(i)\hfill}} \displaystyle \quad \xi= { (c_1+1) x+2\kappa (2c_1-1) t+c_2
\over 3(c_1 t+ c_3)}, \quad \phi={u\over c_1 t+ c_3},\\[3.0ex]
{\hbox to 30pt{(ii)\hfill}} \displaystyle  \quad \xi={x+4 \kappa t + c_1 \over 3t+c_2},
\quad \phi=0. 
\end{array}
$$
These are all equivalent to classical infinitesimals. 	Hence in this case there
are no new nonclassical symmetries.

\subsection{$\epsilon=1$} 

As discussed in the preamble to this section,
we must consider, in addition to the general case of the determining
equations, each of the singular cases of the determining equations. \\

\noindent{\sc Case 3.2.1} {$\xi\not=u$.} 
We can remove factors of ($\xi-u$)
from the determining equations, and we have then that
$\xi_u=0$. Reducing the remaining eight determining equations
\wrt\ this, only the last six are non-zero:
$$
\begin{array}{l}
 3 \phi_{uu} u^{2} - 6 \xi \phi_{uu} u + \beta \phi_{u} u + 3 \xi^{2} \phi_{uu}
 - \beta \xi \phi_{u} - \beta \phi + \beta \xi \xi_{x} + \beta \xi_{t}=0,\\[2.0ex]

\phi_{uuu} u - \xi \phi_{uuu} + \beta \phi_{uu}=0,\\[2.0ex]

\xi_{x} \phi_{u} u - \beta \xi \phi_{x} + \phi \phi_{uu} u + \beta \phi_{x} u
       - \xi \phi \phi_{uu} - 5 \xi \phi_{xu} u+4\xi\xi_{xx}u+\phi_{tu} u \\[0.7ex] 
-\phi \phi_{u} + \xi_{t} \phi_{u} - \xi \phi_{tu} - \xi^{2} \xi_{xx}
    + 3 \phi_{xu} u^{2} - 3 \xi_{xx} u^{2}-2 \xi_{x}^{2}u-2\xi_{xt} u \\[0.7ex] 
+ 2 \xi^{2} \phi_{xu} + 2 \xi_{x} \phi - 2 \xi_{t} \xi_{x} + 2 \xi \xi_{xt} =0,\\[2.0ex]

2 \xi \kappa \phi_{x} - \phi_{t} u + \alpha \phi_{x} u^{2} - 2 \kappa \phi_{x} u
     - \alpha \xi \phi_{x} u 
     + 2 \phi_{xu} \phi_{x} u 
     + \phi \phi_{xxu} u  
     - 2 \xi \phi_{xu} \phi_{x} \\[0.7ex]
     - \xi_{xx} \phi_{x} u 
     - \xi \phi \phi_{xxu} - 3 \xi_{x} \phi u + 2 \xi \xi_{x} \phi 
     + \xi \xi_{xx} \phi_{x} 
     + \xi_{x} \phi_{xx} u 
     - \xi \phi_{xxx} u \\[0.7ex]
     - \xi_{t} \phi + \phi^{2} 
     - \phi \phi_{xx} + \phi_{xxx} u^{2} 
     + \phi_{xxt} u - \xi \phi_{xxt} + \xi_{t} \phi_{xx} + \xi \phi_{t}=0,\\[2.0ex]

2 \beta \phi_{xu} u - \xi_{x} \phi_{uu} u - \beta \xi_{xx} u + 2 \phi_{u} \phi_{uu} u
      + \beta \xi \xi_{xx} - 5 \xi \phi_{xuu} u 
      - \xi \phi \phi_{uuu} \\[0.7ex] 
      + \phi \phi_{uuu} u 
      + \phi_{tuu} u    
      - \phi \phi_{uu} 
      + \xi_{t} \phi_{uu} - \xi \phi_{tuu} 
      + 3 \phi_{xuu} u^{2} 
      + 2 \xi^{2} \phi_{xuu} \\[0.7ex]
      - 2 \xi \phi_{u} \phi_{uu} 
      + 2 \xi \xi_{x} \phi_{uu} 
      - 2 \beta \xi \phi_{xu} =0,\\[2.0ex]

    4 \xi_{x} \kappa u - 2 \phi_{uu} \phi_{x} u - \beta \phi_{xx} u 
  - \xi \xi_{xx} \phi_{u} - 2 \kappa \phi - 2 \phi \phi_{xuu}u
  -2 \phi_{xtu} u
  +\xi\phi \\[0.7ex] 
  +\xi_{xxt} u
  + \xi_{xxx} u^{2} 
  + \xi_{t} \xi_{xx} 
  - 3 \phi_{xxu} u^{2} + 2 \phi \phi_{xu}
  - \xi \xi_{xxt} 
  + 2 \xi^{2} \xi_{x}
  -\xi^{2}\phi_{xxu} \\[0.7ex] 
  -2\xi_{t}\phi_{xu}
  - \alpha \xi_{t} u 
  + \alpha \xi \phi + \xi_{xx} \phi_{u} u - 2 \xi \xi_{x} \kappa
  - \xi \xi_{xxx} u  
  - 3 \xi \xi_{x} u 
  + 2 \xi \phi_{uu} \phi_{x} \\[0.7ex] 
  +\xi \xi_{x} \xi_{xx}
  + 4 \xi \phi_{xxu} u
  + \beta \xi \phi_{xx} 
   + 2 \xi \phi \phi_{xuu} 
   + 2 \xi \phi_{u} \phi_{xu}
   -  2 \xi \xi_{x} \phi_{xu} -\alpha \xi_{x} u^{2}\\[0.7ex] 
   + \alpha \xi \xi_{x} u  
   + 2 \xi \phi_{xtu} 
   + 2 \xi_{t} \kappa
   - \xi_{t} u - \xi_{xx} \phi  
   - 2 \phi_{u} \phi_{xu} u =0.
\end{array}
$$
Reducing the fifth of these equations  \wrt\  the fourth yields
$$ (\beta-3)\left[(u-\xi) \phi_u -\phi +\xi \xi_x+\xi_t\right]=0. $$
If $\beta=3$, then one easily finds via another route that the expression
in the second bracket is necessarily zero. The equation for $\phi$ can be solved
to give
$$ \phi= F(x,t)(u-\xi) +\xi \xi_x+\xi_t. $$
When this is substituted into the remaining equations we
can then take coefficients of powers of $u$ to be zero, and our problem
is then easily solved.  As in the $\epsilon=0$ case discussed in \S3.1 above,
it is quite straightforward to solve the resulting equations. The complete
solution set is\\

\noindent (a), if $\alpha\not=-1$
\begin{equation}
\label{enqarbi}
\begin{array}{l}
{\hbox to 30pt{(i)\hfill}}\displaystyle \qquad \xi=c_1, \quad \phi=0, \\[1.0ex]
{\hbox to 30pt{(ii)\hfill}}\displaystyle  \qquad \xi
     ={2\kappa\over (1+\alpha)}-{c_1\over t+c_2}, \quad 
    \phi={2\kappa-(1+\alpha)u \over (1+\alpha)(t+c_2)},
\end{array}
\end{equation}
(b), if $\alpha=-1$
\begin{equation}
 \xi=c_1 t+c_2, \quad \phi=c_1,\label{enqnnegi} 
\end{equation}
(c), if $\alpha=-1$ and $\kappa=0$
\begin{equation}
\xi=c_1 - {c_3\over t+c_2}, \quad\phi={c_1-u \over t+c_2}, \label{enqnnegii}
\end{equation}
(d), if $\beta=-1$ and $\alpha=0$
\begin{equation}
\xi= c_1 x-2c_1 \kappa t +c_2, \quad
\phi=3c_1 u-2c_1^2 x+4c_1^2\kappa t -2c_1c_2 -2c_1 \kappa,
\quad \beta\not=0.\label{enqneg}
\end{equation}
The infinitesimals (\ref{enqarbi})--(\ref{enqnnegii}) give rise to classical 
reductions, but (\ref{enqneg}) gives the following new nonclassical reduction.\\

\noindent{\bf Reduction 3.1}
If in (\ref{enqneg}), we set $c_1\not=0$ and $c_2=0$, without loss of generality,
then we obtain
$$ u(x,t)=w(z) \exp\left(3c_1 t\right) + c_1 z \exp\left(c_1 t\right)+2\kappa,
\quad z=(x-2\kappa t-2\kappa/c_1)\exp\left(-c_1t\right),$$
where $w(z)$ satisfies
$$ w w''' -w'w'' +c_1 z w' -3c_1 w =0.$$

\noindent{\sc Case 3.2.2} {$\xi=u$, $\beta\not=3$, $\beta\not=1$.} 
We generate five \deqs,
the first of which is $\phi_{uu} =0$.
Thus $\phi$ is a linear function of $u$, and substituting this into the
remaining four \deqs, we take coefficients of powers of $u$ to
be zero.  These equations are easily solved to give
$ \phi=0$ provided that $\kappa=0$ and $\alpha=-1$.
The \isc\ and (\ref{fulleqn})\ are then solved to give the simple exact 
solution
$$ u(x,t) = {x +c_1\over t+c_2},$$
where $c_1$ and $c_2$ are arbitrary constants.\\

\noindent{\sc Case 3.2.3} {$\xi=u$, $\beta=1$.}
We consider here the case $\phi_{uu}\ne0$, since taking $\phi_{uu}=0$ yields
the same solution as in Case 3.2.2 above. In this instance the remaining four
\deqs\ are
$$
\begin{array}{l}
12\kappa- 2 \phi_{xuu} u - 6\alpha u - 6 u - 2 \phi \phi_{uuu} - 3 \phi_{u} \phi_{uu}
     - 4 \phi_{xu} - 2 \phi_{tuu}=0,\\[2.0ex]

\phi_{xu} \phi_{xx} u- \phi_{u} \phi_{xxx} u -\alpha \phi_{u} \phi_{x} u
   - \phi \phi_{xu} u - 2 \phi_{x} \phi_{xx} 
     + \phi \phi_{uu} \phi_{xx} 
\\[0.7ex] 
     + \phi_{tu} \phi_{xx} 
     - 2 \phi_{u} \phi_{xu} \phi_{x}
     + 2 \kappa \phi_{u} \phi_{x} + 2 \phi \phi_{x} - \phi^{2} \phi_{uu}
     - \phi \phi_{u} \phi_{xxu} 
\\[0.7ex]
     - \phi_{xxt} \phi_{u} 
     + \phi_{t} \phi_{u} 
     - \phi \phi_{tu}=0,\\[2.0ex]

\phi_{u} \phi_{xuu} u +4\alpha\phi_{u} u  - \phi \phi_{uu}^{2}
   - \phi_{tu} \phi_{uu} + \phi \phi_{u} \phi_{uuu}+6\phi
  -4\phi_{xx}
\\[0.7ex] 
 + \phi_{tuu} \phi_{u} 
 - 4 \phi_{xxu} u 
 + 4 \phi_{u} u 
 - 2 \phi_{uu} \phi_{x} 
  - 4 \phi \phi_{xuu} + 2 \phi_{u}^{2} \phi_{uu} - 8 \kappa \phi_{u} 
\\[0.7ex]
  -2\alpha \phi 
  - \phi_{xu} \phi_{uu} u 
   - 4 \phi_{xtu}=0,\\[2.0ex]

\phi \phi_{uu} u +\alpha\phi_{xu} u^{2} - 2 \phi_{u} \phi_{uu} \phi_{x}
  - 2 \phi \phi_{u} \phi_{xuu} - 2 \phi_{u} \phi_{xxu} u 
  +\alpha\phi_{tu} u \\[0.7ex] 
  + 2 \phi \phi_{xu} \phi_{uu}
  -2 \kappa\phi\phi_{uu}
  + 2 \phi_{tu} \phi_{xu}
  +\alpha \phi \phi_{uu} u 
  + 2 \phi_{xxx} u 
  - 2 \phi_{x} u
\\[0.7ex] 
  + 2 \phi_{xu}^{2} u 
  - 3 \phi_{u} \phi_{xx} 
  + 2 \phi \phi_{xxu}
  - 2 \phi_{u}^{2} \phi_{xu} 
  - 2 \phi_{xtu} \phi_{u}
  + 4 \phi \phi_{u}
 - 2 \kappa \phi_{tu} 
\\[0.7ex] 
  -\alpha\phi \phi_{u} 
  - 2 \kappa \phi_{xu} u
  + \phi_{tu} u 
  + \phi_{xu} u^{2} 
  + 2 \phi_{xxt} - 2 \phi_{t}=0.
\end{array}
$$
Using the procedures in the package {\tt diffgrob2} with an ordering
designed to eliminate first derivatives with respect to $t$, then
derivatives with respect to $x$, one can obtain several equations for
derivatives of $\phi$ with respect to $u$ only.  One can then continue
to produce lower order and lower degree equations in the $u$-derivatives
of $\phi$, using repeated cross-differentiation and reductions. For
example, the ``Direct Search" procedure in the {\tt diffgrob2} manual,
\cite{\refM} may be used.
This process suffers from expression swell.  No termination of this
process was observed by us within the computer memory available, and the
expressions obtained contained thousands of summands!
One of three results appear likely.  Firstly, the process terminates with the
highest derivative term being $\phi$ itself, yielding
$\phi$ to be a function of $u$ alone (note that $x$ and $t$ do not appear
explicitly in any of the determining equations).  Inserting this into the
determining equations, one must have that
$\phi$ is constant, a contradiction to our standing assumption in this subcase. 
Secondly, the process may terminate with an inconsistency, and
thirdly, the process may terminate but with such a large expression that
the result is useless.\\

\noindent{\sc Case 3.2.4} {$\xi=u$, $\beta=3$, $\phi_u\not=0$.} 
Four determining equations were obtained, the first of which is $\phi_{uu}=0$,
so we substitute $\phi=F(x,t)u+G(x,t)$ into the remaining three
and require $F(x,t)\not=0$. We find that there are no such 
solutions.\\

\noindent{\sc Case 3.2.5} {$\xi=u$, $\beta=3$, $\phi_u=0$ and not both $\kappa$ 
and $\alpha+1$ are zero.} 
One \deq\ was obtained which was
a polynomial in $u$ of degree two whose coefficients are functions
of $x,t$ only, so the coefficients of powers of $u$
must be zero. These equations 
were easily simplified using the procedures in {\tt diffgrob2} to yield,
\begin{eqnarray}
&& \kappa\not=0,\quad \alpha=-1,\quad \phi= 0 , \label{xiuini}\\
&& \kappa\not=0,\quad \alpha=-1,\quad \phi= {-2\kappa \over t+c_1},
   \label{xiuinii} \\[1.0ex]
&&
\begin{array}{@{\hspace{0.0ex}}ll}
   \kappa\hbox{ arbitrary},\quad\alpha\not=-1,
      &\phi=c_1\exp(\zeta)+c_2\exp(-\zeta),\\[1.0ex]
      &\displaystyle \zeta=\mbox{i}\sqrt{\alpha}\left(x-{2\kappa t\over1+\alpha}\right).
\end{array}
    \label{xiuiniii}
\end{eqnarray}

In (\ref{xiuini}) if we solve (\ref{fulleqn})\ and the \isc\ as a system
of equations we find that the only solution is $u(x,t)=c$, a
constant. 

In (\ref{xiuinii}) we can solve (\ref{fulleqn})\ and the \isc\ to give the 
exact (canonical) solution
$$ u(x,t)=  -2\kappa+{x/t},$$
which cannot be realised by any of the previously found reductions,
though it would not appear to be a particularly interesting
solution.  It is interesting to note that performing the {\tt KolRitt}
algorithm of {\tt diffgrob2} on the system comprising the original
equation with the \isc\ led to a simple calculation for $u$.  By contrast,
the usual procedure of solving the \isc\ using the method of
characteristics and inserting the result into the original equation
to obtain the reduction was considerably more difficult due to the
implicit nature of the reduction.

In (\ref{xiuiniii}) we can again solve our problem to yield the
exact (canonical) solution
$$ u(x,t)={-2\kappa \over 1+\alpha} \pm (c_0 +c_1 \mbox{e}^{\zeta}
+c_2 \mbox{e}^{-\zeta})^{1/2},
\qquad \zeta=\mbox{i}\sqrt{\alpha}\left(x- {2\kappa t \over 1+\alpha}\right),$$
which is a special case of the travelling wave reduction 2.5.\\

\noindent{\sc Case 3.2.6} {$\xi=u$, $\beta=3$, $\phi_u=0$, $\kappa=0$, $\alpha=-1$.} 
We are left simply with the determining equation $\phi_{xx}-\phi=0$, which
produces the following \inl,
\begin{equation}
 \phi= g(t) \mbox{e}^x +h(t) \mbox{e}^{-x}, \label{xiuiv} 
\end{equation}
where $g$ and $h$ are \afs. Hence we have to solve
the \isc
\begin{equation}
 u u_x +u_t = g(t) \mbox{e}^x +h(t) \mbox{e}^{-x}. \label{eqIIIxv}
\end{equation}
It is straightforward to show that every solution of this equation is also a
solution of (\ref{fulleqn}).

\section{Nonclassical ($\tau=0$) and Direct Methods}
\setcounter{equation}{0}

\def\eztw{\redIIi}
\def\ezkn{\redIIii}
\def\entw{\redIIIi}
In the canonical case of the nonclassical method 
when $\tau=0$ we set $\xi=1$ without loss
of generality.
We proceed by eliminating $u_x,u_{xx},u_{xxx}$ and $u_{xxt}$
in (\ref{fulleqn})\ using the \isc\ (\ref{invsc})\ which yields
\begin{equation}
\label{delstard}
\begin{array}{l}
u_t -\epsilon \phi \phi_{uu} u_t -\epsilon \phi_{xu} u_t
     -\epsilon \phi_{u}^2 u_t  -\phi_{xx} u -\phi_{u} \phi_x u 
     -\phi^2 \phi_{uu} u 
     -2 \phi \phi_{xu} u \\[1.0ex] 
    -\phi \phi_u^2 u 
    -\alpha\phi u-\beta \phi \phi_x -\epsilon \phi_t \phi_u
  -\beta \phi^2 \phi_u -\epsilon \phi_{xt} 
      -\epsilon \phi \phi_{tu}+2 \kappa \phi =0,
\end{array}
\end{equation}
which involves the infinitesimal $\phi$ that is
to be determined. As in the $\tau\not=0$ case we
apply the classical Lie algorithm to this equation using the
first prolongation $ {\rm pr}^{(1)} {\bf v}$ and eliminate
$u_t$ using (\ref{delstard}). 

The equivalent approach using the direct method of Clarkson and Kruskal
\cite{\refCK} is to consider the ansatz $u=U(x,t,w(t))$ and require that
the  result be \ode\ for $w(t)$; see also \cite{\refPACiii,\refLouii}. 
It is straightforward to show that this yields the equivalent reductions.\\

{\bf Case 4.1} {$\epsilon=0$.} 
The nonclassical method generates a single equation of 25 terms, without any
singular solutions. Since this is difficult to solve explicitly,  we seek
polynomial solutions in $u$.\\

\noindent {\it Ansatz 1}.\quad$\phi=F(x,t)$.\quad In this case we
obtain the following three exact solutions for (\ref{fulleqn}) with $\epsilon=0$: 
\begin{equation}
u(x,t)= \mu_2\left[x-(2\kappa-\beta\mu)t\right]^2 + \mu_0, \label{IVsoli}
\end{equation}
where $\mu_2$ and $\mu_0$ are arbitrary constants, provided that
$\alpha=0$,
\begin{equation}
u(x,t)={(x-2 \kappa t)^3\over 12 t}+ \mu(x-2\kappa t) + \delta t^{1/2},
\label{IVsolii}
\end{equation}
where $\delta$ is an arbitrary constant, provided that
$\alpha=0$ and $\beta=-1$, and
\begin{equation}
u(x,t)=-\,{x-2\kappa t \over \alpha t},\label{IVsoliii}
\end{equation}
provided that $\alpha\not=0$.\\

\hide{\noindent {\it Ansatz 2}.\quad$\phi=F(x,t)u+G(x,t)$. \quad The
coefficients of powers of $u$ set to zero yield three equations, one a third
order \ode\ in $F$, the other two both have both dependent variables present. It
doesn't seem  possible to integrate the \ode\ (in order to simplify the problem),
and there is large expression swell when trying to use {\tt diffgrob2}.}

\noindent {\it Ansatz 2}.\quad$\phi=F(x,t)u^2+G(x,t)u+H(x,t)$. \quad 
In this case we obtain the following three exact solutions for (\ref{fulleqn}) with
$\epsilon=0$;
\begin{equation}
u(x,t)=A\tan\left[\tfr12\sqrt\alpha(x-2\kappa t) \right], \label{IVbsoli}
\end{equation}
where $\mu$ is an arbitrary constant, provided that $\beta=-3$,
\begin{equation}
u(x,t)= A \exp\{\mu(x-2\kappa t)\},\qquad
\mu^2= -\,{\alpha\over1+\beta},\label{IVbsolii}
\end{equation}
provided that $\beta\not=-1$, and 
\begin{equation}
 u(x,t)=A\,\mbox{sech}\{\tfr12\sqrt\alpha(x-2\kappa t)\},\label{IVbsoliii}
\end{equation}
provided that $\beta=-3$.\\

{\bf Case 4.2} {$\epsilon=1$.} In this case the nonclassical method generates
a single equation of 150 terms, which has a singular
solution if and only if
$$ \phi\phi_u +\phi_x -u - {2\kappa/\beta} =0, $$
provided that $ \alpha-\beta-1 =0$.
We again seek polynomial solutions of $\phi$
using one ansatz.\\

\noindent {\it Ansatz 1}.\quad$\phi=F(x,t)$.\quad In this case we
obtain three following three exact solutions for (\ref{fulleqn}) with $\epsilon=0$:
\begin{equation}
u(x,t)= \mu_2\left[x-(2\kappa-\beta\mu)t\right]^2 +
\mu_1\left[x-(2\kappa-\beta\mu)t\right]+\mu_0, \label{IVcsoli}
\end{equation}
where $\mu_2$,
$\mu_1$ and $\mu_0$ are arbitrary constants, provided that $\alpha=0$,
\begin{equation}
\label{IVcsolii}
\begin{array}{r}
\displaystyle
u(x,t)={(x-2 \kappa t)^3\over 12 t}+{\mu_2(x-2 \kappa t)^2\over t}
+ \left({1+8\mu_2^2\over2t}+\mu_1\right)(x-2\kappa t) + \delta t^{1/2}\\[2.5ex]
\displaystyle
+{\mu_2(6+16\mu_2^2)\over3t}+2\kappa + \mu_1+\mu_2,
\end{array}
\end{equation}
where $\mu_2$,
$\mu_1$ and $\delta$ are arbitrary constants, provided that
$\alpha=0$ and $\beta=-1$, and
\begin{equation}
u(x,t)=-\,{x-2\kappa t \over \alpha t},\label{IVcsoliii}
\end{equation}
provided that $\alpha\not=0$.

\section{Discussion}
\setcounter{equation}{0}

In this paper we have classified symmetry reductions of the nonlinear third
order \pde\ (1.1), which contains three special cases that have attracted
considerable interest recently, using the classical Lie method and the
nonclassical method due to Bluman and Cole \cite{\refBC}. The use of the MAPLE
package {\tt diffgrob2} was crucial in this classification procedure. In
the classical case it identified the special cases of the parameters for
which additional symmetries might occur whilst in the nonclassical case, the
use of {\tt diffgrob2} rendered a daunting calculation tractable and thus
solvable. 

In their recent paper, Gilson and Pickering \cite{\refGP} discuss the
application of the Painlev\'e tests for integrability due to Ablowitz, Ramani and
Segur \cite{\refARS,\refARSi} and Weiss, Tabor and Carnevale \cite{\refWTC} to
equation (1.1). In particular, they investigate the integrability of the
\odes\ arising from the travelling-wave reductions 2.1 and 2.5 above. It would
be interesting to investigate the integrability of some of the \odes\ arising
from the other reductions derived in this paper using standard 
Painlev\'e analysis,
``weak 
Painlev\'e analysis'' \cite{\refRDG,\refRRDG} and ``perturbative 
Painlev\'e analysis''
\cite{\refCFP}, though we shall not pursue this further here. Marinakis and
Bountis \cite{\refMB} have also applied 
Painlev\'e analysis to the FFCH equation (\ref{ch});
an interesting aspect of their analysis is the use of a hodograph transformation.
To conclude we remark that the RH equation (\ref{rh}) is a quasilinear
\pde\ of the form discussed by Clarkson, Fokas and Ablowitz \cite{\refCFA}. It
is routine to apply their algorithm, which involves a hodograph transformation,
for applying the 
Painlev\'e PDE test to such quasilinear \pdes\ and show that (\ref{rh})
does not satisfy the necessary conditions to be solvable by inverse
scattering.

\vspace{3.0ex}

\begin{center}
\large \bf Acknowledgments
\end{center}

We thank the editors for inviting us to write an
article. We also thank the Program in Applied Mathematics, University of
Colorado at Boulder, for their hospitality during our visit whilst some of this
work was done. The research of PAC and ELM is supported by EPSRC (grant
GR/H39420) and that of TJP by an EPSRC Postgraduate Research Studentship, which
are gratefully acknowledged.

\def\refbk#1#2#3#4#5  {\bibitem{#1}{\frenchspacing#2},{\frenchspacing\sl#3}, #4\ (#5).}
\def\refpp#1#2#3#4    {\bibitem{#1}{\frenchspacing#2}, #3, #4.}
\def\refjl#1#2#3#4#5#6{\bibitem{#1}{\frenchspacing#2}, 
                                   {\frenchspacing\it#3},\ {\bf#4}, #5\ (#6).}
\def\refeb#1#2#3#4#5#6#7{\bibitem{#1}{\frenchspacing#2}, 
                                     {\frenchspacing\it#3},\ {\bf#4}\ no.#5, #6\ (#7).}
\def\reftoap#1#2#3#4#5{\bibitem{#1}{\frenchspacing#2},
                                   {\frenchspacing\it#3}, #4\ (#5).}

\end{document}